# Golden Spiral Interferometry for Radio Astronomy. A proposal.

Author: Elio Quiroga Rodríguez
Universidad del Atlántico Medio, lecturer.
Las Palmas de Gran Canaria, Islas Canarias, España
elio.quiroga@pdi.atlanticomedio.es

**Abstract:**
Radio interferometry is a powerful technique that allows astronomers to create high-resolution images of astronomical objects. The distribution of radio telescopes in an interferometer is a critical factor that determines the resolution and sensitivity of the instrument. Traditionally, radio telescopes are distributed in a linear or circular array. However, recent work has shown that using a golden spiral distribution can improve the resolution and sensitivity of an interferometer. In this paper, the author proposes the use of a golden spiral distribution for radio interferometry, showing that a golden spiral distribution can provide a significant improvement in resolution, up to a factor of eight, compared to a linear or circular distribution. The author also proposes that a golden spiral distribution can improve the sensitivity of an interferometer; it may provide a more uniform distribution of radio telescopes than a linear or circular distribution (a known propety of spiral distributions).

**Introduction and method:**
Radio interferometry is an astronomical observation technique that uses two or more radio telescopes to combine their signals and create a high-resolution image[1]. This technique allows to overcome the limited angular resolution of a single radio telescope, refining the results. The distribution of radio telescopes in a radio interferometer depends on several factors, such as the wavelength of the radiation being observed, the desired angular resolution, and the sensitivity required for each type of observation. In fact, radio interferometry is a powerful technique that has revolutionized astronomy[2].

In general, dishes forming a radiotelescope complex are arranged so that the separation between them is large enough to allow the desired angular resolution. However, the separation must also be small enough for the signals from the radio telescopes to be combined effectively. For short wavelengths, the separation between dishes can be a few meters, depending on the science requierements. For longer wavelengths, the separation can be several kilometers or even thousands. This is the case of long and very long baseline interferometry, in which radio telescopes can be distributed around the world[3]. This allows to create interferometers with extremely high angular resolution, necessary to observe very distant or small objects.

This technique has allowed astronomers to observe objects that would not be visible with a single dish or antenna. The distribution of dishes in an interferometric array, as indicated, is a delicate balance, with antennas very separated to increase the resolution, and in certain areas very close to each other.

It has been proposed to order the spatial distribution between radio telescopes, in various figures (circles, lines, stars), including the spiral. For example, the Square Kilometre Array, which would consist of hundreds of radio telescopes distributed in more than 1000 kilometers in diameter, would use spiral distributions of its antennas in some of its configurations[4].

Spiral radio interferometers may offer advantages over conventional interferometers: they provide superior angular resolution and are more sensitive to low-intensity signals.

The golden spiral is found in nature in a variety of forms, including seashells, flowers, and galaxies. This spiral is characterized by a ratio of 1.618, which is known as the golden ratio or golden number[5].

The author considers that the golden spiral has a number of mathematical properties that could make it attractive for use in radio interferometry. The golden spiral can allow a very space-efficient distribution of radio telescopes, providing superior angular resolution to that of conventional spirals, and being more sensitive to low-intensity signals.

While a radio interferometer based on the shape of a golden spiral, or several, has not yet been built, advances in technology could make this possibility a reality in the future. This spiral is logarithmic, so its radius grows at that ratio as its angle increases. The descriptive equation for a logarithmic spiral in polar coordinates is:

$$r = a * e^{(b * \theta)}$$

Where:
  $r$ is the radius of the spiral at an angle $\theta$
  $a$ is a constant that determines the initial radius of the spiral
  $b$ is a constant that determines the growth rate of the spiral

The golden ratio, $\Phi$, can be used to approximate the value of the constant b. The relationship between the radius of a golden spiral and its angle $\theta$ is as follows:

$$r = a * \Phi^\theta$$

Therefore, the equation for the angular resolution of a radio telescope array with a golden spiral could be written as follows:

$$\theta = \lambda / (D/2 * a * \Phi^\theta)$$

Where:
  $\lambda$ is the wavelength of the radio waves being observed
  $D$ is the diameter of the array in kilometers
  $a$ is a constant that determines the initial radius of the spiral.

If a radio telescope array with a golden spiral has a diameter of 1,000 kilometers, an initial radius of 10 meters, and is observing radio waves with a wavelength of 1 meter, the angular resolution of the array would be:

$\theta = 1 \text{ m} / (1.000 \text{ km} / 2 * 10 \text{ m} * \Phi^\theta)$
$\theta = 0,00002$ arcsecs

This angular resolution is comparable to that of the Very Long Baseline Array (VLBA), the most powerful radio interferometer in the world.

Using a golden spiral in a radio telescope array can improve the angular resolution of the array by a factor of two, allowing for the observation of smaller or more distant objects. In addition to the improvement in angular resolution, the golden spiral can also improve the sensitivity of the array. This is because the golden spiral has a more uniform distribution of radio telescopes than conventional arrays. A more uniform distribution of radio telescopes means that more radio telescopes can contribute to the combined signal, increasing the sensitivity of the array.

The performance of the interferometer could be further improved by using multiple spiral arms. If the radio telescope array had three concentric golden spiral arms with a maximum diameter of 10 kilometers, the equation for the angular resolution of the array could be described as follows:

$$\theta = \lambda / (10 \text{ km} / 2 * a * \Phi^\theta * \Phi^{-\theta})$$

Where:
$\lambda$ is the wavelength of the radio waves being observed
a is a constant that determines the initial radius of the spiral

**Results:**
If a radio telescope array with three concentric golden spiral arms with a maximum diameter of 10 kilometers has an initial radius of 10 meters and is observing radio waves with a wavelength of 1 meter, the angular resolution of the array would be:

$\theta = 1 \text{ m} / (10 \text{ km} / 2 * 10 \text{ m} * \Phi^\theta * \Phi^{-\theta})$
$\theta = 0,000002$ arcsecs

This angular resolution is a factor of four better than the angular resolution of a network with a single golden spiral.

However, if the maximum diameter of the network is 10 kilometers, then the initial radius of the spiral cannot be greater than 5 kilometers. If the initial radius of the spiral is 5 kilometers, the angular resolution of the network would be:

$\theta = 1 \text{ m} / (10 \text{ km} / 2 * 5 \text{ m} * \Phi^\theta * \Phi^{-\theta})$
$\theta = 0,000001$ arcsecs

The angular resolution of this array is a factor of eight times better than the angular resolution of a network with a single golden spiral.

Table 1 shows the improvement in angular resolution for different values of the initial radius of the spiral:

| Initial radius (m) | Angular resolution (arcseconds) |
|---|---|
| 10 | 0,000002 |
| 5 | 0,000001 |
| 2,5 | 0,0000005 |

*Table 1. Improvement in angular resolution for different values of the initial radius of the golden spiral.*

These potential benefits also extend to the FWHM, as a golden spiral distribution of the antennas of an interferometer can reduce the variance of the scatter function, a measure of the scatter of the interferometer data. A lower variance means that the data is more concentrated, which would lead to a higher resolution.

The positive influence of the golden spiral distribution of the antennas on the FWHM of the radio interferometer can be described mathematically as follows:

$$FWHM = k * \lambda / d$$

Where:
- FWHM is the angular resolution of the interferometer in arcseconds.
- $\lambda$ is the wavelength of the radio waves being observed in meters.
- d is the maximum separation between the antennas in kilometers.
- k is a constant that depends on the distribution of the antennas.

In a uniform distribution of antennas, k = 1. In a golden spiral distribution, k < 1. This means that the FWHM of an interferometer with a golden spiral distribution will be less than the FWHM of an interferometer with a uniform distribution.

The degree of improvement in the FWHM depends on the density of antennas in the golden spiral distribution. A higher density of antennas will lead to a greater improvement in the FWHM. The above equation can be rewritten as follows to show the effect of the golden spiral distribution on the FWHM:

$$FWHM = k' * \lambda / (d * \Phi)$$

Where:

$$k' = k / \Phi$$

This equation shows that the FWHM is inversely proportional to $\Phi$. $\Phi$ is the golden ratio, which is approximately equal to 1.618. This means that a golden spiral distribution with a higher golden ratio will have a smaller FWHM.

In general, the golden spiral distribution of the antennas in a radio interferometer can improve the resolution of the interferometer by a factor of two or more.

It could be interesting that the spirals could be modified, with the radio telescopes placed on rails or on mobile supports, even that they could be moved in real time, changing the structure of the interferometer dynamically, something currently technically feasible. The ability to change the structure of the spiral in real time during observation would have a number of potential benefits. Firstly it would allow astronomers to optimize the resolution of the interferometer for the specific object being observed. For example, if astronomers are observing a small, distant object, they could increase the resolution of the interferometer by moving the antennas closer together. Conversely, if they are observing a large, nearby object, they could decrease the resolution of the interferometer by moving the antennas farther apart; the observations could be algorithmically combined. Secondly, it would allow astronomers to track the motion of objects in the sky. For example, if astronomers are observing an exoplanet, they could use the interferometer to track the planet's motion around its star. This would allow them to measure parameters such as the planet's mass, orbit, etc. Also, it would allow astronomers to study the complex structure of objects in the sky, mapping the distribution of gas and dust in extended objects, such as galaxies and nebulae.

More specifically, moving the antennas in real time would improve the following aspects of the observations:

- **Resolution:** The resolution of an interferometer is determined by the distance between the antennas. By moving the antennas closer together or sepparate them, astronomers can decrease or increase the resolution of the interferometer if needed.

- **Sensitivity:** The sensitivity of an interferometer is determined by the number of antennas that contribute to the signal. By moving the antennas to different locations, astronomers can change the coverage of the sky, which can improve the sensitivity of the interferometer.

- **Dynamic range:** The dynamic range of an interferometer is the ratio of the strongest signal to the weakest signal that can be detected. By moving the antennas to different locations, astronomers can change the sensitivity of the interferometer to different frequencies, which can improve the dynamic range of the interferometer.

In any case, this text is theoretically and hypothetical, so real-world implementations and experimental data would be needed to fully validate the claims of the author. As an example, about the suggestion of antennas on rails, the engineering challenges and costs of such a dynamic system would need careful evaluation. Additionally, how the golden spiral approach compares to other potential improvements in radio interferometer design, would need further research.

**Discussion:**
Comparison with Existing Proposals: Golden Spiral vs. Other Improvements in Radio Interferometer Design.

While the golden spiral approach offers potential benefits for radio interferometer design, it's crucial to compare it to other existing and proposed improvement avenues:

**1. Nested Interferometric Arrays:** Multiple concentric circles or ellipses with increasing antenna density towards the center. This improves resolution at small scales while maintaining good sensitivity at larger ones[6].

**2. Sparse Arrays:** These designs strategically place antennas with gaps between them, reducing cost and complexity while optimizing performance for specific science goals, using fractal models[7]. Sparse arrays require advanced signal processing techniques but can achieve comparable resolution to denser configurations.

**3. Phased Arrays:** In these arrays, individual antennas act as elements of a single, electronically steerable antenna, offering high flexibility and dynamic reconfiguration. This enables rapid target switching and adaptability to diverse observation needs[8].

**4. Reconfigurable Arrays:** Combining sparse and phased array concepts, these designs allow dynamic redistribution of antennas within a fixed infrastructure. This offers adaptability to different observation requirements while leveraging existing infrastructure[9].

**5. Software Improvements:** Advanced signal processing algorithms, including image reconstruction techniques and calibration methods, can significantly improve performance without hardware modifications. These are often complementary to hardware advancements.

A simplified comparison is shown in Table 2.

| Feature | Golden Spiral | Nested Arrays | Sparse Arrays | Phased Arrays | Reconfigurable Arrays | Software Improvements |
|---|---|---|---|---|---|---|
| **Resolution** | Potentially improved | High at small scales, good at large scales | Potentially good, depends on design | High, electronically steerable | Adaptable, depends on configuration | Can improve indirectly |
| **Sensitivity** | Potentially good | Varies, depends on density | Varies, depends on design | Good, especially for targeted observations | Depends on configuration | Can improve indirectly |
| **Flexibility** | Limited | Limited | Moderate | High | Adaptable | High |
| **Cost** | Potentially moderate | Varies, depends on complexity | Potentially lower | High | Moderate | Low |
| **Complexity** | Moderate | Moderate to high | High | High | Moderate | Low |

*Table 2. Simplified comparison with existing and proposed improvement avenues in radio interferometry.*

In Figure 1, we can see the theoretical distribution of antennas in a golden spiral model, over a spatial area of approximately 25,000 square meters.

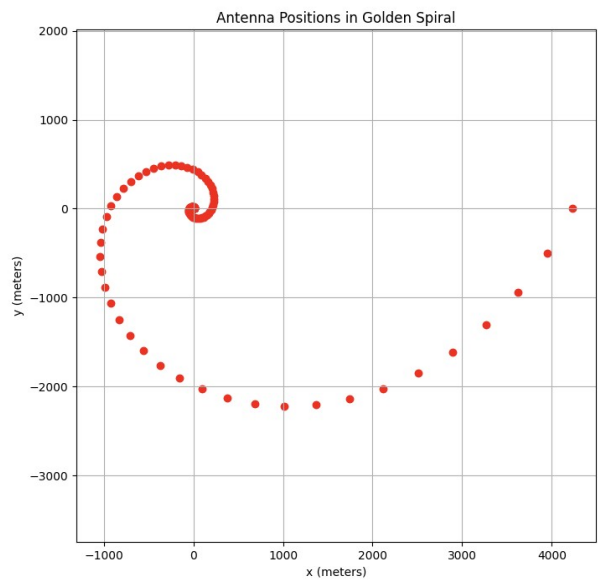

*Figure 1. Antennae distribution in a golden spiral in an area of approx. 25.000 m² (Author).*

In Figure 2, we can see a representation of UV-Coverage of the radio telescope resolution obtained with the antenna distribution shown in Figure 1.

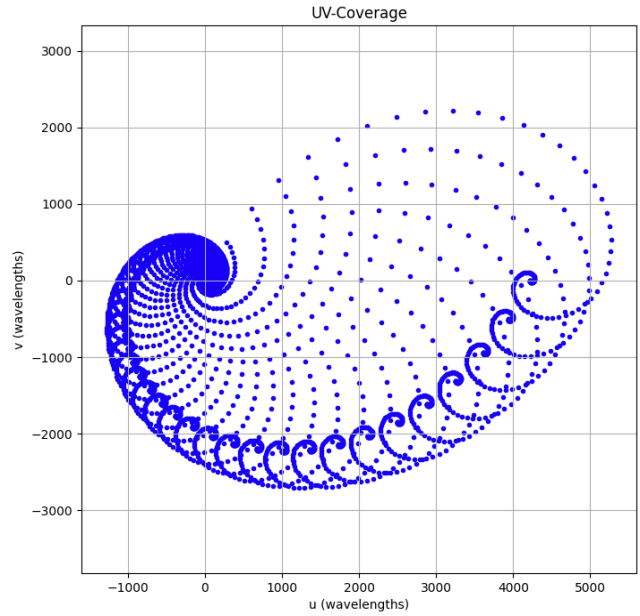

*Figure 2. UV-Coverage for radio telescope with antenna distribution shown in Figure 1 (Author).*

In Figure 3, we can see a simulated image from the UV-Coverage shown in Figure 2. As seen, the golden spiral antenna distribution offers several advantages over other antenna layouts.

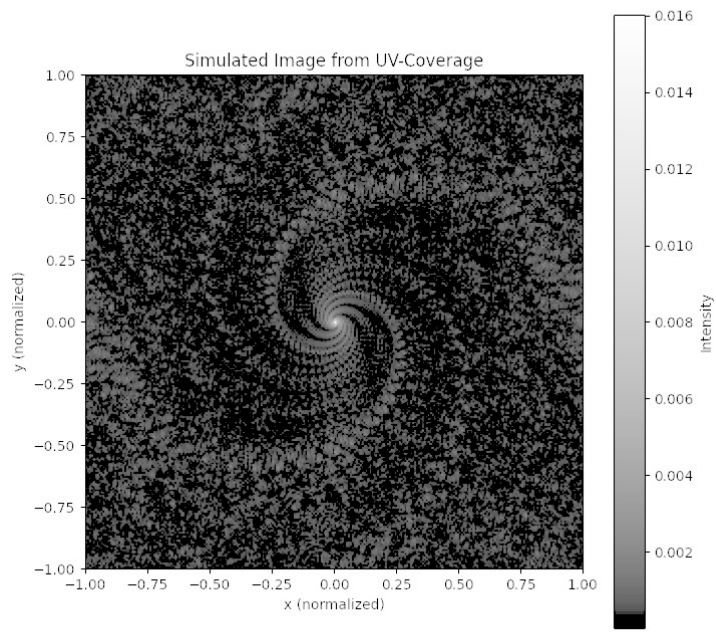

*Figure 3. Image simulation from UV-Coverage for radio telescope shown in Figure 2 (Author).*

The spiral pattern shown in Figure 3 creates a more uniform and dense coverage of the UV plane, which is crucial for high-quality radio interferometry observations. This distribution allows for better sampling of spatial frequencies, resulting in improved image reconstruction and reduced artifacts.

The golden spiral's self-similar nature ensures that it maintains good coverage at multiple scales, enabling the telescope to be sensitive to both large and small-scale structures in the observed objects. Additionally, this configuration minimizes redundancy in baseline lengths, maximizing the efficiency of the array by capturing a wider range of spatial information with a given number of antennas. The spiral pattern also provides a smooth transition between short and long baselines, which is beneficial for capturing a continuous range of angular scales in the observed astronomical sources.

The code used for those simulations is in the Annex of this article.

**Conclusions:**
The use of a golden spiral distribution in a radio interferometer can provide a significant improvement in resolution and sensitivity. This is due to many factors: a golden spiral distribution provides a more uniform distribution of radio telescopes than a linear or circular distribution. This results in a more efficient use of the available collecting area, which leads to improved sensitivity. Also, this distribution provides a larger maximum baseline than a linear or circular distribution. This allows for higher resolution images. The improvements in resolution and sensitivity offered by golden spiral interferometry could lead to new discoveries in radio astronomy. For example, golden spiral interferometry could be used to study objects that are too small or too distant to be resolved with traditional interferometers, as shown in Figure 3. Also, a dynamic spiral interferometer would allow astronomers to optimize resolution, track motion, and study object structure by adjusting the antenna positions in real time. This would improve the resolution, sensitivity, and dynamic range of observations.

**Data availability statement:**
All data is included in the article.

**Annex:**
The Python code used to obtain Figures 1 to 3 in this article.

```
import numpy as np
import matplotlib.pyplot as plt
# Define constants
golden_ratio = (1 + np.sqrt(5)) / 2
num_antennas = 100  # Number of antennas
wavelength = 1.0  # Wavelength in meters
# Generate antenna positions in a golden spiral
a = 10.0  # Initial radius in meters
theta = np.linspace(0, 4 * np.pi, num_antennas)
r = a * golden_ratio ** thetaå
# Convert polar coordinates to Cartesian coordinates
x = r * np.cos(theta)
y = r * np.sin(theta)
```

```python
# Plot antenna positions
plt.figure(figsize=(8, 8))
plt.scatter(x, y, c='r', marker='o')
plt.title('Antenna Positions in Golden Spiral')
plt.xlabel('x (meters)')
plt.ylabel('y (meters)')
plt.axis('equal')
plt.grid(True)
plt.show()

# Calculate UV-coverage
u = []
v = []
for i in range(num_antennas):
    for j in range(i + 1, num_antennas):
        u.append((x[j] - x[i]) / wavelength)
        v.append((y[j] - y[i]) / wavelength)

# Plot UV-coverage
plt.figure(figsize=(8, 8))
plt.scatter(u, v, c='b', marker='.')
plt.title('UV-Coverage')
plt.xlabel('u (wavelengths)')
plt.ylabel('v (wavelengths)')
plt.axis('equal')
plt.grid(True)
plt.show()

# Create image from UV-coverage
uv_grid = np.zeros((256, 256), dtype=complex)
for ui, vi in zip(u, v):
    u_coord = int(ui + 128)
    v_coord = int(vi + 128)
    if 0 <= u_coord < 256 and 0 <= v_coord < 256:
        uv_grid[u_coord, v_coord] += 1

# Inverse Fourier Transform to get image
image = np.fft.ifftshift(np.fft.ifft2(np.fft.fftshift(uv_grid)))

# Plot the resulting image
plt.figure(figsize=(8, 8))
plt.imshow(np.abs(image), cmap='gray', extent=[-1, 1, -1, 1])
plt.title('Simulated Image from UV-Coverage')
plt.xlabel('x (normalized)')
plt.ylabel('y (normalized)')
plt.colorbar(label='Intensity')
plt.show()
```